\documentclass[%
 reprint,
 amsmath,amssymb,
 aps,
]{revtex4-2}

\usepackage{float}
\usepackage[utf8]{inputenc}
\usepackage{natbib}
\usepackage{graphicx}
\usepackage{dcolumn}
\usepackage{bm}
\usepackage{xcolor}
\usepackage{soul}
\usepackage{hyperref}

\begin{document}

\preprint{APS/123-QED}

\title{Quantifying nematic order in evaporation-driven self-assembly of Halloysite nanotubes: Nematic islands and critical aspect ratio}

\author{Arun Dadwal$^{1}$}
\author{Meenu Prasher$^{2}$}\email{meenu@barc.gov.in}
\author{Pranesh Sengupta$^{2,3}$}
\author{Nitin Kumar$^{1}$}
\email{nkumar@iitb.ac.in}

\affiliation{
	$^1$Department of Physics, Indian Institute of Technology Bombay Powai, Mumbai 400076, India. \\
 $^2$Materials Science Division, Bhabha Atomic Research Centre, Mumbai 400085, India \& \\
	 $^3$Homi Bhabha National Institute, Anushakti Nagar, Mumbai 400094, India \\
	}
\date{\today}






\begin{abstract}
Halloysite nanotubes (HNTs) are naturally occurring clay minerals found in Earth's crust that typically exist in the form of high aspect-ratio nanometers-long rods. Here, we investigate the evaporation-driven self-assembly process of HNTs and show that a highly polydisperse collection of HNTs self-sort into a spatially inhomogeneous structure, displaying a systematic variation in the resulting nematic order. Through detailed quantification using nematic order parameter $S$ and nematic correlation functions, we show the existence of well-defined isotropic-nematic transitions in the emerging structures. We also show that the onset of these transitions gives rise to the formation of \textit{nematic islands} $\textendash$ phase coexisting ordered nematic domains surrounded by isotropic phase $\textendash$ which grow in size with $S$. Detailed image analysis indicates a strong correlation between local $S$ and the local aspect ratio, $L/D$, with nematic order possible only for rods with $L/D \ge 6.5 \pm 1$. Finally, we conclude that observed phenomena directly result from aspect ratio-based sorting in our system. Altogether, our results provide a unique method of tuning the local microscopic structure in self-assembled HNTs using $L/D$ as an external parameter.

\end{abstract}

\maketitle


\section{Introduction}
Clay minerals refer to fine-grained hydrous layered silicate minerals, known as phyllosilicates, found abundantly in the Earth's crust. They consist of $[\text{Si}_2\text{O}_5]^{2-}$ tetrahedral layer (T) bonded to an Al-OH/Mg-OH octahedral layer (O) to form T-O sheets arranged in layers with inter-layer spacing 0.7 nm-1.5 nm \cite{BERGAYA20131,kumari2021basics, BRIGATTI200619}. Halloysite  $(\text{Al}_2\text{Si}_2\text{O}_5(\text{OH})_4\cdot n\text{H}_2\text{O})$; n=0 (dehydrated), 2 (hydrated)) is one such naturally occurring clay \cite{wilson2016global}, composed of 15-20 T-O sheets rolling to give them unique nanoscroll morphology known as Halloysite Nanotubes (HNTs) \cite{yuan2008functionalization,yuan2015properties,joussein2005halloysite}. At the microscopic level, they appear as rods with a high aspect ratio. The HNTs have interesting properties, including thermally stable hydroxyl groups \cite{ASGAR2021126106}, an empty lumen \cite{D1EN01032H}, and bio-compatibility \cite{Hanifdrug-delivery, GODA20191083}, which make them suitable for a variety of applications \cite{MAHMOUD20231585,vergaro2010cytocompatibility,massaro2017halloysite,danyliuk2020halloysite,lisuzzo2022pickering, ZHANG20168}. Achieving these requires better control and understanding of their self-assembled structures at the microscopic length scales.

A dense collection of high aspect ratio rods forming a lyotropic nematic liquid crystalline phase is a universal phenomenon observed in many systems across a vast spectrum of length scales\cite{wijnhoven2005sedimentation,narayan2007long,mourad2009sol,xu2011aqueous,puech2011highly,liu2018liquid,nystrom2018confinement, bagnani2018amyloid, kadar2021cellulose,kumar2022catapulting}. Such a phase of matter is characterized by a high degree of orientational order, which can be quantified using a director field \cite{de1993physics} representing the average local orientation of the rods. Recently, aqueous dispersions of HNTs have been shown to exhibit a lyotropic phase transition between isotropic and nematic states with profound implications on their physical properties \cite{luo2013liquid}. However, a detailed characterization of the nematic order in the aqueous medium poses challenges, preventing a comprehensive understanding of the complex behavior of these systems. Thus, an alternate approach is required, providing an easy yet extensive way to quantify emerging nematic order in these systems.

Self-assembly of nanoscopic particles using the evaporative technique provides one such method to investigate complex pattern formation at microscopic scales \cite{song2003nematic, zhang2010ordering,ming2008ordered,nobile2009self,lin2011self,dugyala2013shape,dugyala2015evaporation,li2016evaporative,zaibudeen2021understanding,khawas2023anisotropic,almohammadi2023evaporation}. Consequently, in the past few years, many such studies have been performed with HNTs, revealing the presence of emerging nematic order and topological defects \cite{zhao2015orientation, liu2017self}. However, most of these studies were performed at lower concentrations ($c$ $<$ 5 wt. \%) with the nematic phase confined to a narrow region near the trailing edge of the drop \cite{zhao2015orientation,liu2021tunable}. Whether or not more interesting structures form at higher $c$, and a general approach to quantify the emerging nematic order in terms of the order parameter, $S$, is lacking. 

In this study, we investigate the emergence of nematic order in evaporation-driven structures of HNTs. We dry out an aqueous droplet containing charge-stabilized HNTs with a high degree of polydispersity in the rod length ($L$) and diameter ($D$) at $c$ = 20 wt. \%. Later, we take multiple SEM images of the dried droplet captured from the center to the edge at regular intervals. We perform comprehensive image analysis to extract the nematic director field and measure the nematic order parameter, $S$, as a function of radial distance. We find that the rods self-assemble into three regions: isotropic near the center, nematic at a distance of about half the droplet radius, and, again isotropic near the droplet edge. In addition, we also report hitherto unknown structures near phase transition regions resembling highly ordered local nematic domains surrounded by an isotropic phase, which we call nematic islands. The size of these islands grows with the nematic order and closely matches the nematic correlation length. In all our experiments, we find a strong linear correlation between local $S$ and the local rod aspect ratio, $L/D$, signifying size-dependent sorting. Finally, our data suggests that a highly ordered nematic structure is possible only for $L/D \ge 6.5 \pm 1$. Our findings reveal a universal protocol for tunable self-assembled structures in nanorod systems, with $L/D$ as a convenient control parameter.

\section{Results $\textbf{\&}$ Discussion}

Pristine HNTs contain impurities such as quartz and illite \cite{huang2023sodium}. We remove these impurities through a purification process, as shown in Fig. 1 (a). This process involves treating the pristine HNTs with sodium hexametaphosphate [(NaPO$_{3}$)$_{6}$] in alkaline pH, leading to the purification and augmentation of their surface charge (see experimental methods section and procedure described in \cite{luo2013liquid}). Fig. 1(b) compares zeta potential, $\zeta$, between the pristine HNTs and purified HNTs, which shows an increase from -16.7 mV to -36.6 mV after purification. This increase is caused by excess deprotonated silanol groups (Si-OH) on the outer surface of purified HNTs \cite{zeng2014facile}. The higher negative $\zeta$ value enhances colloidal stability, facilitating their self-assembly process. Additionally, we perform TEM imaging to characterize the shape and dimensions of purified HNTs, as shown in Fig. 1 (c). The TEM micrograph reveals the tubular shape and empty lumen with a 12-20 nm diameter. We also capture SEM images of HNTs (Fig. 1(d)) to determine the length ($L$) and outer diameter ($D$), as shown in Fig. 1(e). We find average values $\langle L \rangle$, $\langle D \rangle$, and $\langle L/D \rangle$ of HNTs to be 0.62 $\pm$ 0.38 $\mu$m , 93 $\pm$ 21 nm, and 7.8 $\pm$ 4.5, respectively. The distributions are quite broad, suggesting a high degree of polydispersity in our system.

\begin{figure}
	\centerline{\includegraphics[width=0.5\textwidth]{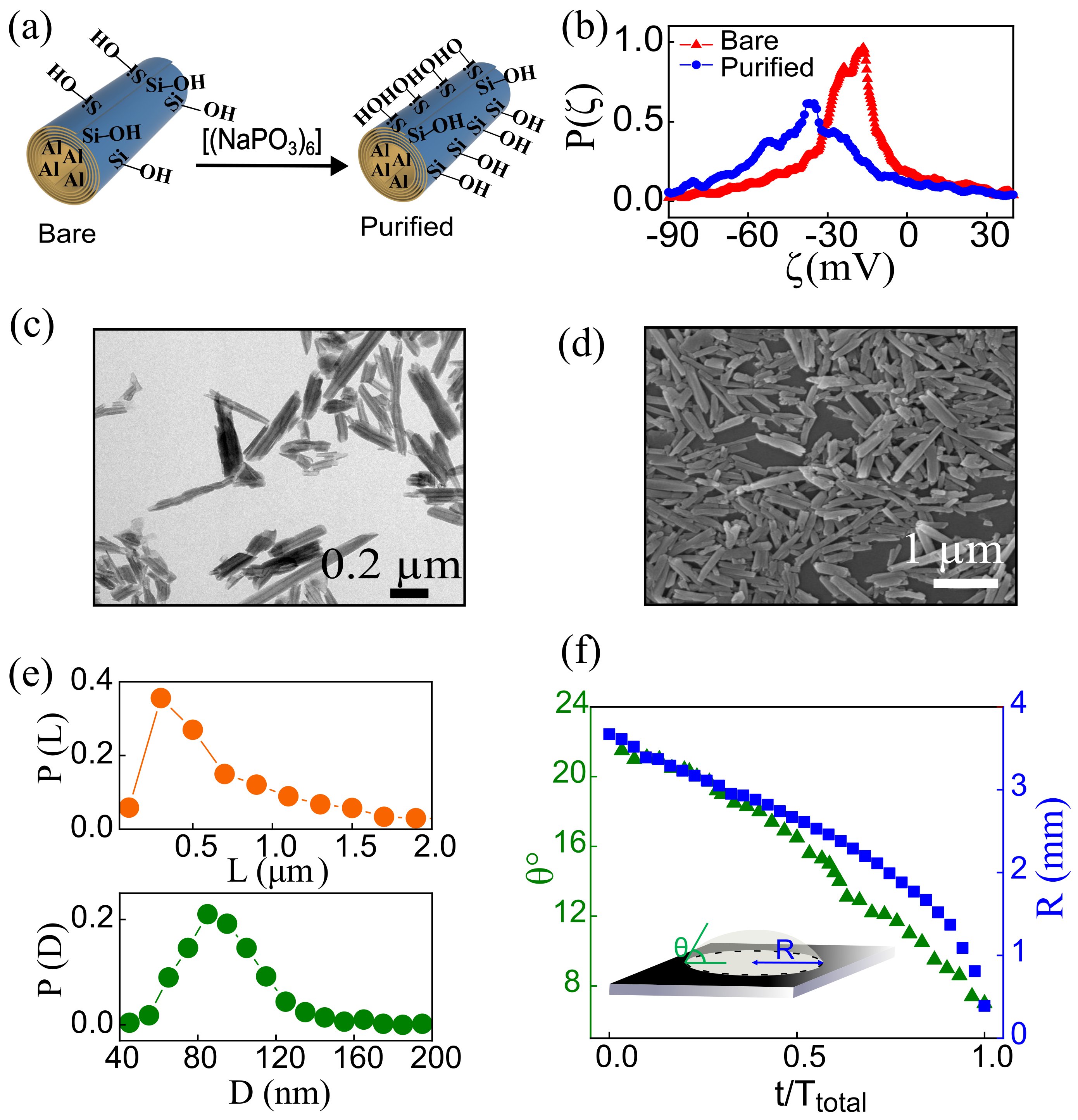}}
	\caption{Characterisation of purified HNTs: (a) Schematic illustration of the purification process of HNTs. This process involves treating the bare rods with sodium hexametaphosphate, Na(PO$_{3}$)$_{6}$, in alkaline pH, leading to the purification and augmentation of their surface charge. (b) The resulting $\zeta$ potential measurement shows an enhanced increase in the average $\zeta$ of the purified rods. (c\&d) TEM and SEM micrographs of HNTs used in the experiment show an empty lumen and irregularly shaped tips. (e) The length ($L$) and diameter ($D$) distributions of the HNTs used in the experiment. (f) Time evolution profile for contact radius R and contact angle $\theta$ with dimensionless time, $t/T_{total}$, where $T_{total}$ = 25 min is the total evaporation time.}
	\label{Figure-1}
\end{figure}

\begin{figure*}
	\centerline{\includegraphics[width=0.8\textwidth]{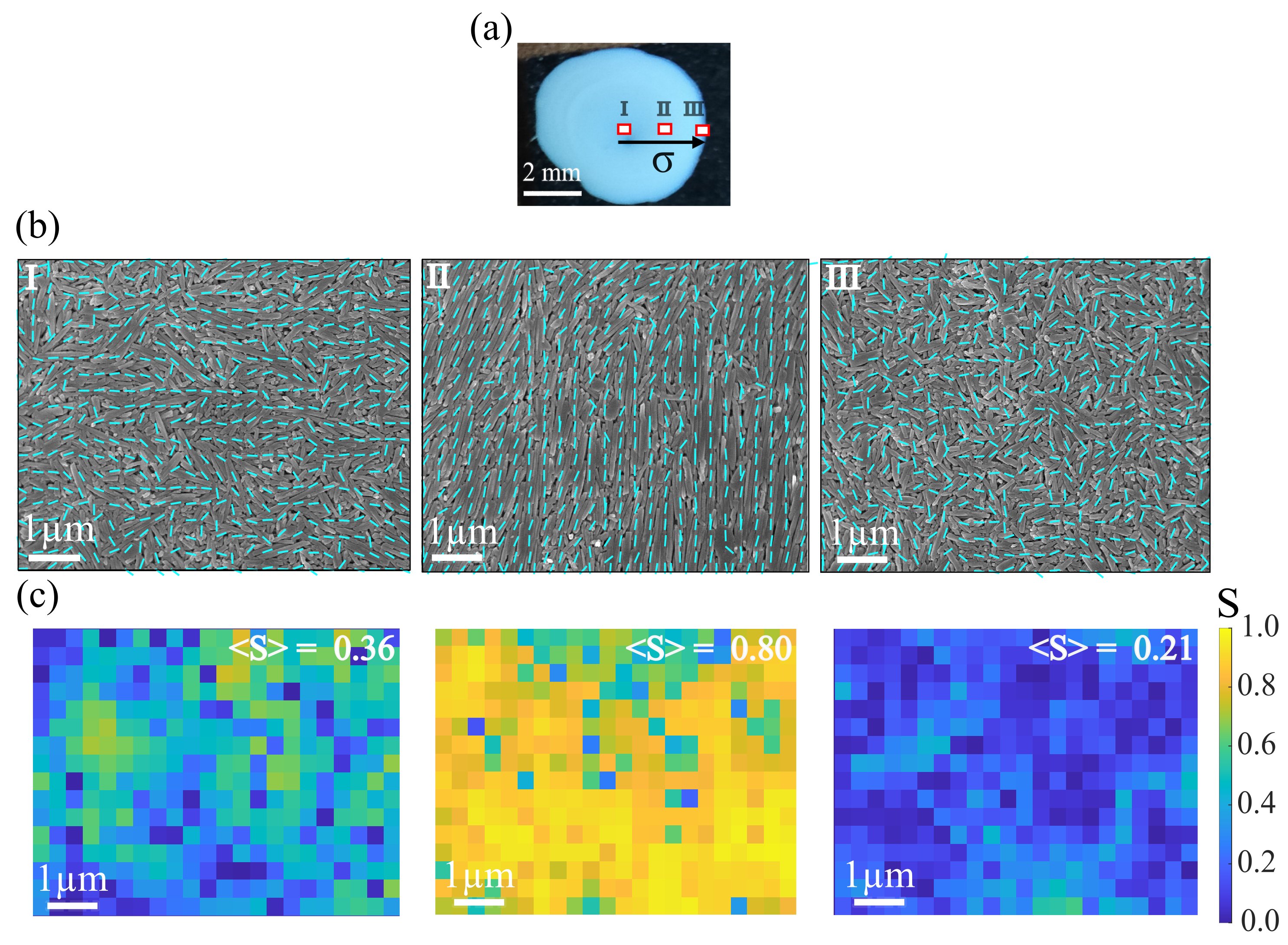}}
	\caption{The emergence of nematic order in the dried droplet of HNTs (a) A typical photograph of a dried droplet on the silicon substrate. We take multiple SEM images along the radial direction as a function of $\sigma$ (distance in the unit of droplet radius, $r/R_d$), with three typical images marked as I, II $\&$ III. (b) SEM images of the marked positions in (a) with cyan lines superimposed on the images indicating the director field. (c) The plot of corresponding local nematic order, $S$, with $\langle S\rangle$ representing its averaged value over the entire image.}
	\label{Figure-2}
\end{figure*}

To gain insights into the self-assembly process, we first analyze the evaporation dynamics of sessile droplets, the results of which are summarized in Fig. 1 (f). We begin by drying a droplet of volume $V$ = 10 \textmu{}L containing HNTs at a concentration $c$ = 20 wt.\% on a hydrophilic silicon substrate under ambient conditions. Fig 1(f) shows a typical droplet drying profile, which suggests a continuously receding droplet radius, $R$, ensuring a weak coffee-ring effect in our system \cite{almohammadi2023evaporation}. Moreover, the contact angle, $\theta$, also decreases with time, indicating that our system undergoes a mixed drying mode \cite{xu2013evaporation,parsa2018mechanisms}. Therefore, we expect a combination of high $c$, $V$, and mixed drying modes to lead to a uniform deposition throughout the surface.

To quantify the underlying order in the self-assembled HNTs, we capture SEM images at three different locations in a radially outward direction, as shown in Fig. 2(a). The position of the image is identified by the parameter $\sigma$, which is the radial distance $r$ in the units of droplet radius, $R_d$ (i.e., $\sigma = r/R_d$, $R_d$ $\approx$ 3 mm). Three typical SEM images of the marked positions in Fig. 2(a) are shown in Fig. 2(b). The cyan lines represent the director field generated using an algorithm originally developed by Cetera et al. \cite{cetera2014epithelial}, which captures the underlying orientation of HNTs quite faithfully.  Here, an individual director is evaluated by averaging the local direction of rods inside a square-shaped area of $\mathcal A$ = 0.06 $\mu$m$^2$ (see Image analysis in the Experimental Methods section). We use this director-field information to extract the local two-dimensional nematic order parameter, $S= 2\langle \cos^2 \theta \rangle-1$, where $\theta$ denotes the angle subtended by a rod with its average, $\langle \theta \rangle$, calculated over all the rods inside $\mathcal A$. $\langle \cos^2 \theta \rangle$ is averaged over all the rods in that region. The value of $S$ ranges between 0 and 1, corresponding to a randomly oriented and perfectly aligned state, respectively. The results are shown as a heat map in Fig. 2(c), where the color represents the local $S$ value with its average value, $\langle S \rangle$, mentioned in each plot. Clearly, $\langle S \rangle$ seems to vary non-monotonically with $\sigma$. To confirm this, we systematically take multiple images from the droplet's center to the edge and calculate $\langle S\rangle$ for each image as a function of $\sigma$, as shown in Fig. 3(a). The plot shows a sustained period of low $\langle S\rangle$ before sharply rising to a higher value at $\sigma \approx 0.45$, which persists up to $\sigma \approx 0.75$ before dropping again abruptly. These results indicate that the system first undergoes an isotropic-nematic (I $\textendash$ N) and then a nematic-isotropic (N $\textendash$ I) transition with increasing $\sigma$.

\begin{figure*}
	\centerline{\includegraphics[width=1\textwidth]{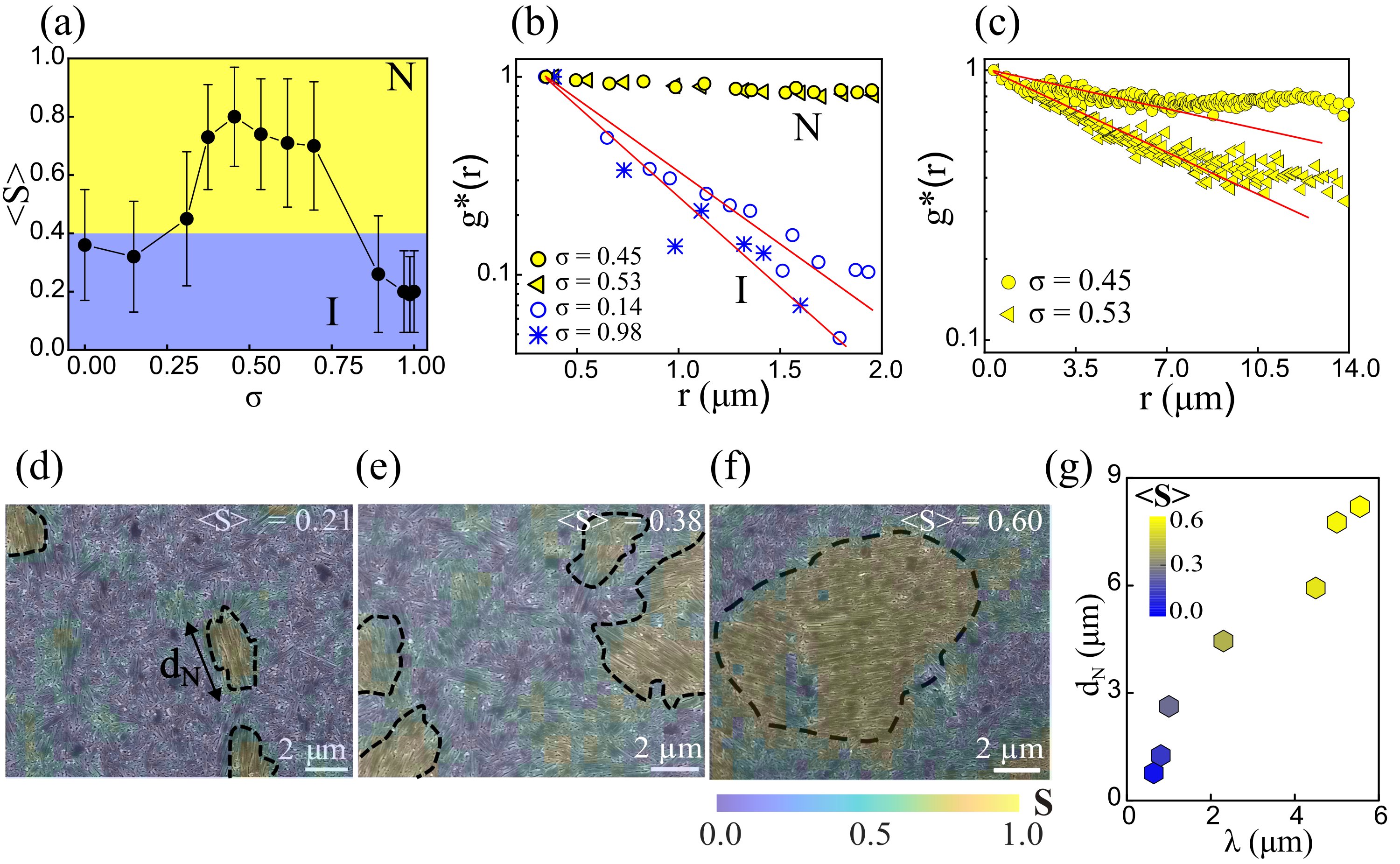}}
	\caption{Observation and characterization of Isotropic-Nematic transitions and resulting nematic island formation in the self-assembled structures of HNT's: (a) Variation of $\langle S\rangle$  as the function of $\sigma$. Yellow and blue regions indicate nematic (N) and isotropic (I). (b) The normalized nematic correlation functions, $g^*(r)$ in a semi-log plot. Yellow curves (indicated by symbol N) show a prolonged decay, whereas blue curves (indicated by symbol I) decay much faster to zero. Solid lines indicate the exponential fit ($g^*(r) = e^{-r/\lambda}$), providing a correlation length $\lambda$ (c) An expanded view of $g^*(r)$ in the nematic state, showing an initial exponential decay (solid lines) followed by a saturation to a finite value. (d-f) SEM images superimposed with local $S$ show the presence of nematic islands (outlined with black dashed lines). $d_N$ is the average island size. (g) $d_N$ increases linearly with $\lambda$, indicating that correlation length determines the
island size. The increase in $d_N$ $\&$ $\lambda$ also ensues the onset of the nematic state in the system measured in terms of $\langle S \rangle$.}
	\label{fig3}
\end{figure*}

This is further validated by evaluating the nematic correlation function, $g_2(r) = \langle\cos[2(\theta_i-\theta_j)]\rangle_{ij}$, which measures the propensity for directors $i$ $\&$ $j$ averaged over all pairs separated by a distance $r$ to orient in the same direction. We calculate its normalized value defined as $g^*(r) = g_2(r)/g_2(r_{min})$ (i.e., scaled by its value at the smallest $r$), as shown in Fig. 3(b). Clearly, $g^*(r)$ can be divided into two groups. The first one, indicated by yellow symbols, shows a slow decay with correlation persisting over large $r$. Its expanded view in Fig. 3(c) even shows an initial exponential decay followed by a saturation to a finite value indicative of a true long-range order. On the other hand, the second group (blue symbols in Fig. 3(b)) exhibits a much faster exponential decay to zero, suggestive of a disordered phase. It is worth mentioning that in Fig. 3(c), we do not see a signature of quasi-long-range order in the form of power-law decaying correlation functions, as one might expect in a two-dimensional system \cite{Chaikin1995, Kosterlitz1973, Chat2006, Shankar2018}. 
Regardless, we reckon these two groups correspond to nematic and isotropic phases, thus pointing to a first-order phase transition in our system.

Next, we fit exponential functions of the form $g^*(r) = e^{-r/\lambda}$ to all the curves and compute a nematic correlation length $\lambda$. Based on the resulting $\lambda$ values, we divide $\langle S\rangle$ vs. $\sigma$ in Fig. 3(a) into well-defined isotropic (I) and nematic (N) regions with $\langle S \rangle = 0.4$ marking the transition boundary. Therefore, we conclude that the rods undergo two discontinuous phase transitions with increasing $\sigma$: I $\textendash$ N and then N $\textendash$ I. Surprisingly, even though these phase transitions happen under non-equilibrium conditions driven by a complex evaporation process, they still show similar properties to an equilibrium isotropic-nematic transition for hard spherocylinders reported in the past \cite{Bolhuis1997}. 

Furthermore, we also see signatures of these transitions in the form of phase-coexistence. Figs. 3(d), (e) $\&$ (f) captured for various $\langle S\rangle$ across the transition line clearly show co-existing nematic and isotropic phases. We observe highly ordered nematic regions (yellow, marked by a dashed line) surrounded by an isotropic background (blue), where the colors indicate local $S$. We name them \textit{nematic islands}, which grow in size as a function of $S$. The observation of these islands is similar to the previously reported phenomenon of size fractionation where longer rods tend to form distinct nematic structures in a two-sized binary mixture \cite{Hamade2020,donaldwindlehanna2006, Birshtein1988}. We measure the average island size, $d_N$ (estimated by taking the square root of the area and taking the average over all visible islands), and find that it shows a near-linear correlation with $\lambda$ as shown in Fig. 3(g). This implies that the correlation length determines the island size near the transition regions. The symbol colors represent the value of $\langle S\rangle$, suggesting that both $\lambda$ and $d_N$ increase with the degree of nematic order. In conclusion, the onset of nematic order leads to increasing nematic correlation length, which provides a way to determine the size of nematic islands in the system.

\begin{figure*}
	\centerline{\includegraphics[width=0.8\textwidth]{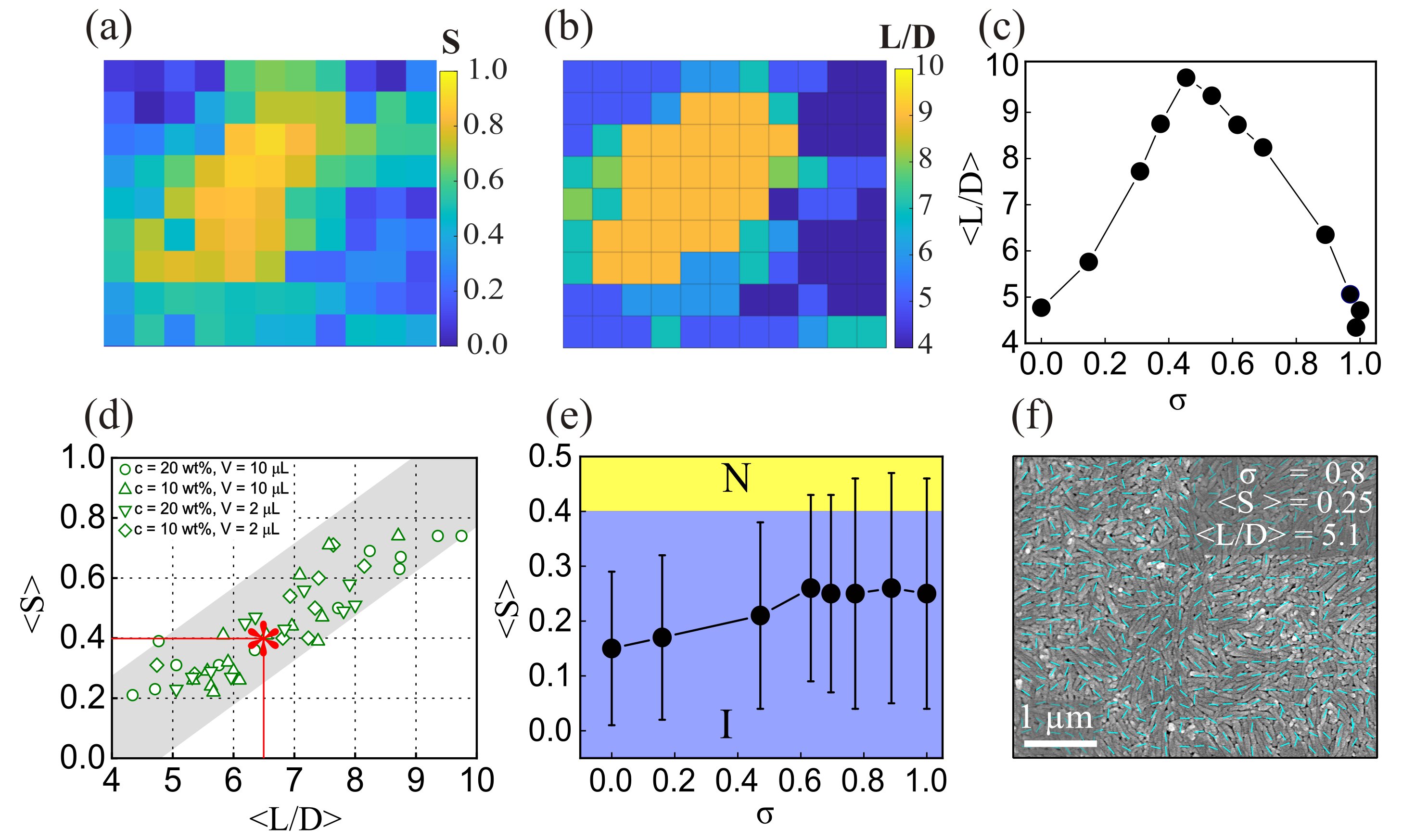}}
	\caption{Rod aspect ratio determines the degree of nematic order in self-assembled structures: (a) Heat map displaying the spatial variation of $S$ for the SEM image shown in Fig. 3(f). (b) Heat map representing the local distribution of $L/D$ visually matching with (a). (c) Variation of $\langle L/D\rangle$ as a function of $\sigma$ corresponding to the data presented in Fig. 3(a) showing a similar non-monotonic dependence. (d) A linear correlation between the $\langle S \rangle$ and $\langle L/D \rangle$ plotted for various $c$ \& $V$ implying $L/D$ as governing parameter for local nematic order. The red asterisk (\textcolor{red}{$\ast$}) corresponds to the critical value of $L/D$ required to observe the nematic state (= 6.5 $\pm$ 1). (e) Variation of $\langle S \rangle$ with $\sigma$ for a sample prepared at $\langle L/D \rangle$ = 5.1. The phase remains isotropic, with $\langle S \rangle$ never crossing into the nematic phase phase. (f) A typical isotropic phase is seen in the SEM image for $\sigma$ = 0.8 for the sample shown in (e).}
	\label{fig4}
\end{figure*}

Next, we aim to uncover the parameter that determines the nematic order in self-assembled structures. To this end, we plot heat maps of $S$ and $L/D$, shown in Figs. 4(a) $\&$ (b) respectively, corresponding to the SEM image of Fig. 3(f). Interestingly, we find a strong spatial correlation between them down to the micrometer length scales. Later, we measure the average aspect ratio $\langle L/D\rangle$ as a function of $\sigma$, where the averaging is performed over all the rods visible in the image. Notably, the $\langle L/D\rangle$ vs. $\sigma$ curve (Fig. 4(c)) shows qualitatively similar non-monotonic dependence as in Fig. 3(a). This corroborates our claim that $S$ and $L/D$ are correlated with each other. To further test the robustness of this result, we perform experiments with droplets of varying $V$ and $c$ and plot their $\langle S\rangle$ vs. $\langle L/D\rangle$ together in Fig. 4(d). We find that the linear relationship is valid for all the experiments over a wide range of $\langle S \rangle$ and $\langle L/D\rangle$. This implies that the aspect ratio alone plays a crucial role in determining the local nematic order in self-assembled HNTs. Using $\langle S \rangle = 0.4$ as the I-N phase transition point (Fig. 3(a)), we estimate critical $L/D = 6.5 \pm 1$ (indicated by a red asterisk symbol in Fig. 4(d)) beyond which the nematic phase appears in the system.

To put this number into perspective, previous numerical studies show a critical $L/D$ of 7 for two-dimensional cases on a square lattice in strictly equilibrium conditions \cite{Ghosh2007, Bates2000}, which drops down to 6 when non-equilibrium effects are considered \cite{Patra2018}. An even lower $L/D$ of 5 has been reported for numerical studies conducted in continuum limit for three dimensions \cite{Frenkel1988, Bolhuis1997}, and in the monolayer of vibrated granular rods \cite{Narayan2006}. Considering the complexity of self-assembly in our experiments, it is hard to determine similarities and differences with previous studies. Nevertheless, our findings demonstrate that an $ L/D$-dependent self-sorting is happening in our system, which, in turn, determines the local nematic order. This result underscores the importance of rod-anisotropy in affecting self-assembled structures.

Finally, we put this hypothesis to the test by specifically preparing a modified HNT dispersion with an $\langle L/D\rangle$ = 5.1 ± 2.0 (see methods), much below the observed critical value of 6.5 $\pm$ 1. In this case, $\langle S\rangle$ increases monotonically with $\sigma$ but never crosses into the nematic state, as shown in Fig. 4(e). A typical SEM image with the director field superimposed shown in Fig. 4(f) verifies that the rods always exist in the isotropic phase even at the highest $\sigma$. These findings confirm our premise that the aspect ratio alone is sufficient to tune the order of HNTs.

\section{Conclusions}

Here, we explored the evaporation-based self-assembly to investigate liquid crystalline behavior observed within a sessile droplet of HNTs dispersion. We performed these experiments with much higher volume fraction, aspect ratio polydispersity, and droplet volume using a mixed mode of drying so that we get a uniform deposition of rods throughout the surface. Our observations revealed that the polydispersity of rods within the droplet led to self-sorting behavior, forming a spatially inhomogeneous nematic state resulting in well-defined Isotropic-Nematic transitions. We observed the formation of novel structures, called nematic islands, signifying phase co-existence in our system. Their size increased linearly with nematic correlation length and the nematic order parameter value. We also found that the local aspect ratio of the HNTs solely determined the local nematic order parameter, which was true for all samples irrespective of the initial drop volume and concentration of the HNTs. From our data, we found a critical aspect ratio of $6.5 \pm 1$ required to observe an orientationally ordered structure of rods. These findings highlight the tunability and robustness of the underlying order of the assembly with aspect ratio as a control parameter. 

The formation of phase-coexistence structures in the context of evaporation dynamics is not strange. Evaporation of liquid droplets containing a polydisperse collection of amyloid fibers has shown tactoids in the evaporating liquid \cite{almohammadi2023evaporation}. Tactoids are distinct droplets of nematically aligned rods in an isotropic background. It will be interesting to see whether such structures also form in our system owing to our size polydispersity and if there is any connection between tactoids and the islands reported here.

Finally, the reported phenomena are a delicate interplay of a high degree of polydispersity in rod aspect ratio and an aspect ratio-dependent isotropic-nematic phase transition. Since the dynamics of the evaporation and precipitation process are physical in nature, we believe that our results are not unique to HNTs but should be valid universally, irrespective of the chemical composition. That is to say, any charge-stabilized colloidal suspension of a given size polydispersity, concentration, and volume should give rise to similar microstructures. Thus, using HNTs as a model system, our study provides valuable insights into the emergence of nematic structures in the evaporation-driven assembly of anisotropic nanorods.

\section*{Acknowledgements}
NK acknowledges financial support from DST-SERB for CRG grant number CRG/2020/002925 and IITB for the seed grant. MP acknowledges the financial support from BARC for the departmental project "Irradiation induced transformations in nuclear materials" (UID: RBA4043). MP thanks Dr. Rajib Ganguly, Chemistry Division, BARC, for Zeta potential measurements. AD gratefully acknowledges IIT Bombay for providing a post-doctoral fellowship. AD thanks Dr. Sunita Srivastava for permitting the use of her laboratory facilities and Somnath Paramanick for his assistance with the figures. NK thanks Dr. Guruswamy Kumaraswamy, Dr. Dibyendu Das and Dr. Sriram Ramaswamy for their insightful comments. We thank SAIF-IITB for the TEM facility and Dr. Rejin for his help with the FEG-SEM (Dept. of Chemical Engineering) facility.

\section{Experimental methods}
\subsection{Materials}
Halloysite nanotubes (HNTs) and sodium hexametaphosphate [(NaPO$_{3}$)$_{6}$]  were procured from Sigma Aldrich, while sodium hydroxide (NaOH) was obtained from Merck Chemicals. For the assembly studies, single-side polished Si wafers of $\langle$100$\rangle$ orientation were used. The 1 cm × 1 cm substrates were used after cleaning with freshly prepared piranha solution. The cleaning process effectively removed the organic residue and imparted hydrophilicity to the substrate.

\subsection{Purification of HNT\lowercase{s}}

HNTs were purified using the method reported in the literature \cite{luo2013liquid}. In a typical procedure, 25 g of halloysite powder was added to a flask containing 100 mL of water, and the resulting mixture was stirred continuously for two hours. Na(PO$_{3}$)$_{6}$ (2 g) was gradually added to the above mixture under continuous stirring, followed by adjusting the pH of the resulting mixture to a range of 8-9 using a 10 wt.\% aqueous NaOH solution. Subsequently, this mixture was stirred at room temperature for 24 hours and left standing for two hours to precipitate the aggregates. The impurities and larger HNTs aggregates precipitated at the flask's bottom while individual HNTs remained in the supernatant. The supernatant was collected and then centrifuged at 3000 rpm for 5 min. The supernatant was decanted again, followed by centrifugation at 9000 rpm for 10 min. The resulting precipitates were washed multiple times using deionized (DI) water until they became neutral. Finally, the obtained solid was dried at room temperature.

\subsection{Tuning the aspect ratio of HNTs}
This was achieved by slightly modifying our functionalization protocol which is as follows. Firstly, we performed centrifugation at 3000 rpm to separate the supernatant, which was then subjected to subsequent centrifugation at 6000 rpm. The supernatant was then separated again and subjected to centrifugation at 10,000 rpm. The resulting solid material thus obtained was then dried at room temperature.

\subsection{Droplet deposit and assembly Characterization}

The deposited HNTs were investigated using a field emission gun (FEG)-scanning electron microscope (JSM 7600F) operating at an acceleration voltage of 5 keV. The morphology of HNTs was observed using a transmission electron microscope (TEM) JEOL model JEM 2100F operated at an accelerating voltage of 200 keV. The Zeta potential of the nanoparticle suspensions was measured using Anton Paar Litesizer 500 particle analyzer by electrophoretic light scattering method. Contact angle measurements were carried out using a video camera equipped with a macro lens.

\subsection{Image Analysis}
\textbf{Director field analysis:} Using ImageJ software \cite{abramoff2004image}, we first band-pass filtered the SEM image to remove pixel noise and achieve a uniform brightness across the image. Later, we used the unsharp mask algorithm to make the rod edges more pronounced. The resulting images were analyzed using an algorithm described in the methods of Cetera et al. \cite{cetera2014epithelial}. This algorithm relies on a 2D Fast Fourier transform on a small local section ({area $\mathcal A$}), giving a vector orthogonal to the actual director field. Depending upon the resolution and the clarity of the original SEM image, we vary $\mathcal A$ anywhere between $0.05-0.2$ $\mu m^2$. Smaller $\mathcal A$ values provide much finer details of the nematic director but are prone to errors. Greater values give a more accurate representation but tend to lose finer details in the structure. However, our results are statistically independent of the choice of $\mathcal A$.

\bibliography{references}

\end{document}